\documentclass{iaus}
\usepackage{graphicx,pstricks}
\newcommand{\mconv}{{\cal M}_{conv}}

\title[IAUS 271.~~Astrophysical Dynamics: From Galaxies to Stars] %% give here short title %%
{Time-dependent Turbulence in Stars}

\author[W. David Arnett \& Casey Meakin]   %% give here short author list %%
{W. David Arnett$^1$
 \and Casey Meakin$^1$}

\affiliation{$^1$Steward Observatory, University of Arizona, \\ 
Tucson AZ 85721,  USA \\ email: {\tt wdarnett@gmail.com} \\[\affilskip]
}

\pubyear{2011}
\volume{271}  %% insert here IAU Symposium No.
\pagerange{121--128}
\jname{Astrophysical Dynamics: From Galaxies to Stars}
\editors{Nic Brummell \& Sacha Brun, eds.}
\begin{document}

\maketitle

\begin{abstract}
Three-dimensional (3D) hydrodynamic simulations of shell oxygen burning by \cite{ma07b} 
exhibit bursty, recurrent fluctuations in turbulent kinetic energy. These are shown to be due to
a global  instability in the convective region, which has been suppressed in 
simulations of stellar evolution which use mixing-length theory (MLT). 
Quantitatively similar behavior occurs in the model of a convective roll (cell)
 of \cite{lorenz}, which is known to have a strange attractor that gives rise to 
 random fluctuations in time. 
An extension of the Lorenz model, which includes Kolmogorov damping and
nuclear burning, is shown to exhibit bursty, recurrent fluctuations like those seen in
the 3D simulations. 
A simple model of a convective layer (composed of multiple Lorenz cells)
gives  luminosity fluctuations
which are suggestive of irregular variables (red giants and supergiants, see
\cite{ms75}. Details and additional discussion may be found in \cite{am11}.

Apparent inconsistencies between \cite{amy09} and \cite{nsa} on the nature of
convective driving have been resolved, and are discussed.

\keywords{turbulence, irregular variables, convection}
\end{abstract}

\firstsection % if your document starts with a section,
\section{Introduction}

Three-dimensional fluid dynamic simulations of turbulent convection in an
oxygen-burning shell of a presupernova star show bursty fluctuations which
are not seen in one-dimensional stellar evolutionary calculations (which use
various versions of mixing-length theory, MLT, \cite{bv58}). 

Our particular example is a set of simulations of oxygen burning in a shell of
a star of $23 \rm M_\odot$ (\cite{ma07b,ma10}. This is of astronomical interest 
in its own right as a model for a supernova progenitor, but
also happens to represent a relatively simple and computationally efficient case,
and has general implications for the convection process in all stars.

Three-dimensional hydrodynamic simulations of shell oxygen burning
exhibit bursty, recurrent fluctuations in turbulent kinetic energy (see Fig.~1, left panel).
These simulations show a damping, and eventual cessation, of turbulent motion 
if we artificially turn off the nuclear burning (\cite{amy09}). Further investigation
by \cite{ma10}
shows that nearly identical pulsations are obtained with a volumetric energy generation rate
which is constant in time, so that {\em the cause of the pulsation is independent of any
temperature or composition dependence in the oxygen burning rate.}
Heating is necessary to drive the convection; even with this  time-independent rate 
of heating,  we still get pulses in the turbulent kinetic energy.
Such behavior is fundamentally different from traditional nuclear-energized 
pulsations dealt with in the literature (e.g., the $\varepsilon$-mechanism, \cite{ledoux41,ledoux58,unno89}),
and is a consequence of time-dependent turbulent convection (it might be called
a "$\tau$-mechanism", with $\tau$ standing for turbulence).

\section{A Controversy Resolved.}
This section was written with the aid of extensive
email discussions with A. Nordlund and R. Stein.
\cite{amy09}, analyzing the 3D simulations of \cite{ma07b}, found that buoyancy
driving is  balanced by viscous (Kolmogorov) dissipation; the remaining terms
(mostly the kinetic energy flux escaping the convection zone)
were less than 6\% of the total. \cite{nsa}
found that the buoyancy terms exactly cancelled, and that the viscous dissipation
was balanced by average gas pressure work. Which is correct? Upon careful
reanalysis, both are, as we shall show. 

\subsection{The Convective Mach Number.}
There are two limiting cases for convective flow, depending upon the convective
Mach number $\mconv$ (the ratio of the fluid speed to the local sound 
speed). The case $\mconv \ll 1$ corresponds to  ``incompressible" flow.
For turbulent motion, the pressure perturbation $P'$ is related to the convective Mach
number by $P'/P \sim \rho u_{rms}^2/P  \sim \mconv^2$, and must be small.
Sound waves outstrip fluid motion, so that pressure differences quickly become small, except
possibly for a static background stratification. Most of the historical research on convection
(e.g., the B\`enard problem, \cite{ch61,ll59}) is done in this limit 
(the Bousinesq approximation, \cite{ch61}).

Let $\langle a \rangle$ denote the average
of any variable $a$ over a spherical shell (lagrangian stellar coordinate).
The density perturbation $\rho'$ and the velocity perturbation $u'$ are both first
order in $\mconv$, while the pressure terms (both gas and turbulent pressure)
and the average velocity $\langle u \rangle$ are of second order. At low $\mconv$ the first order
terms dominate, but as $\mconv$ rises, the second and higher order terms become
important, changing the physical behavior of the system. $\mconv$ does not increase
indefinitely for a quasi-static system; as it approaches unity, kinetic energy approaches
internal energy in magnitude, and the system becomes gravitationally unbound,
allowing no quasi-static solution.
There is a narrow range, $ 0.1 \leq \mconv \leq 1$, in which this interesting transition
occurs. Except in dynamic situations, convection in stellar interiors satisfies $\mconv \leq 0.1$,
for which the Boussinesq limit is a reasonable approximation. Near stellar surfaces of
cool stars, $\mconv \geq 0.1$, and the Boussinesq limit is no longer accurate.

\subsection{Convective Driving.}
We are interested in the rate of
transfer of energy into different forms, so we consider various terms for power per unit
volume, or energy per unit volume per unit time. 
In order to fix the terminology, let us define some terms:
\begin{itemize}
\item {\em buoyancy power} is $W_B = \langle \rho' {\bf u' \cdot g}\rangle$, 
\item {\em gas pressure power} is $W_P = \langle -{\bf u  \cdot \nabla} P \rangle$, 
\item {\em net rate of work done by gravity} is $W_G = \langle \rho {\bf u \cdot g}  \rangle$,
\item{\em net rate of work done by gravity on the mean flow} is 
$W_{Gm}= \rho_0 {\bf u_0 \cdot g}$, and
\item {\em net rate of gas pressure work on the mean flow} is 
$W_{Pm} = -{\bf u_0 \cdot \nabla }P_0 $.
 \end{itemize}
For a quasi-steady state, $\langle \partial \rho / \partial t \rangle = 0$ so that
$\langle {\bf \nabla \cdot} \rho {\bf u} \rangle = 0$,
which requires
$0 = \langle \rho u_z \rangle = 
\langle \rho \rangle \langle u_z \rangle + \langle \rho' u_z' \rangle$. 
This is true for all values of $\mconv$. This implies that the total work done by gravity vanishes, but can be split into a (positive) work done on the convective motions, called buoyancy power,
and a (negative) work done on the mean flow, that is, $W_G =W_B + W_{Gm}=0$.
Ignoring boundary and wave effects for the moment, the gas pressure power is balanced
by viscous (Kolmogorov) dissipation, $W_P = \varepsilon_K$, see \cite{ma07b}.

\subsection{Rotational Flow.}
Historically $W_B$ has been used in the Boussinesq approximation as the driving
term for convection (\cite{ch61}).  Using the Cowling approximation we may
move the gravity outside the average, so that 
 $W_B = g_z \langle \rho'u_z'\rangle = -g_z \langle \rho \rangle \langle u_z \rangle $.
 In the Boussinesq limit,  the equation of hydrostatic equilibrium holds,
 $dP/dz = - \rho g_z$, and 
 the rate of work done on the fluid passing through the pressure gradient is 
 \begin{equation}
 \langle -{\bf u \cdot \nabla}P\rangle \rightarrow g_z \langle \rho u_z \rangle = W_B.\nonumber
 \end{equation}
The mathematical result is the same as if the buoyancy power $\cal B$ alone were used, and
provides consistency with the literature (e.g., \cite{ch61,ll59,davidson}). 
Thus, the general result that, integrated over a convection zone, the buoyancy work
is balanced by the viscous dissipation, is still valid in the incompressible limit (\cite{ch61}).
The same result is found for turbulent flow by \cite{amy09}: buoyancy power is balanced
by Kolmogorov dissipation.

For the simulations of \cite{ma07b}, $\mconv \le 0.03$ so that the pressure
fluctuations are small ($P'/P \le 0.001$), clearly in the incompressible regime. In this limit
\cite{amy09} show that $\langle \rho' u'/ \rho \rangle \propto \langle T' u' /T\rangle$, so that the
the convective velocity field is directly related to the enthalpy flux. This connection
is ignored in MLT, but is important for stellar evolution because {\em it removes the freedom
to adjust} the MLT parameter $\alpha$. 

The flow is accelerated by a torque in the horizontal plane, and tends to be
divergence free (solenoidal, or ``rotational") because of mass conservation. 
The velocities are small
enough so that, to the same level of approximation, ram pressure may be ignored, and
the background stratification is hydrostatic.

\subsection{Divergent Flow.}

The ``compressible'' limit is $\mconv \simeq 1$, for which $P'/P \sim 1$. 
Shock formation is the most startling
change in the flow character. Sound wave generation increases rapidly as 
$\mconv \rightarrow 1$ (\cite{ll59}, \S75). The flow becomes diverging, or
``irrotational'' (consider
the extreme limit of a point explosion which is pure divergence). Even in the case of
convective flow, ram pressure levitation begins to become important (\cite{sn98}).
Generally the flow begins to cause structural change and becomes even more complex.
In this case the simple picture of pure buoyancy driving begins to break down; increasingly
more work is done to generate diverging flows as $\mconv$ increases.

If the convective region does not have an overall divergence (i.e., is in a quasi-steady state),
the excess pressure fluctuations (beyond those implied by the buoyancy flux) must
generate waves (both sound waves and gravity waves), which may propagate 
beyond the convective region. The ram pressure
from the flow will cause the quasi-steady convection zone to expand as $\mconv$ increases.
Because of turbulence, episodes of vigorous local pulsation may occur.
Further increase in $\mconv$ may lead to vigorous global pulsation and even explosion 
($\mconv \sim 1$ implies the kinetic energy is of order the internal energy).
Notice the transition from rotational flow toward diverging flow as $\mconv$ increases.

\subsection{Relevance to Astrophysics}
Which of the $\mconv$ limits is relevant for astrophysics? Both are. 
{\em Almost all} the matter in stellar
convection zones, during {\em almost all} evolution, is in the limit of incompressible flow.
For the Sun, the region for which $\mconv \geq 0.1$ (so that the pressure fluctuations are
greater than $0.01$), contains only about $10^{-7}$ of a solar mass.
However all stars with cool surfaces have a superadiabatic region in which the convective
Mach numbers rise to $\mconv \sim 0.1$ or more, and {\em these regions are the ones observed.}
Quasi-static oxygen burning, one of the most vigorous nuclear processes to drive convection,
has $\mconv \ll 1$, and this is true for other thermonuclear
processes as well; but they evolve to violent and explosive events for which $\mconv \sim 1$.
The exceptions to $\mconv \ll1$ are important: (1) explosions, such as supernovae and novae,
(2) vigorous thermonuclear flashes, (3) vigorous pulsations, especially radial ones,
and (4) the sub-photospheric layers of stellar surface convection zones, which are
strongly non-adiabatic, to name a few. 
Consequently the transition (near $ \mconv \sim 0.3$, implying pressure perturbations of
order $0.1$) 
between these two limiting cases is emerging as an important problem in astrophysics.

We have made some theoretical progress in understanding the $\mconv \ll 1$ case, and
will now follow Eddington's advice to ``push a theory until it breaks," to see what happens.

\section{Turbulence and the Lorenz Model}

\begin{figure}
\mbox{
\includegraphics[trim=0 -10 30 35,angle=0,scale=0.4]{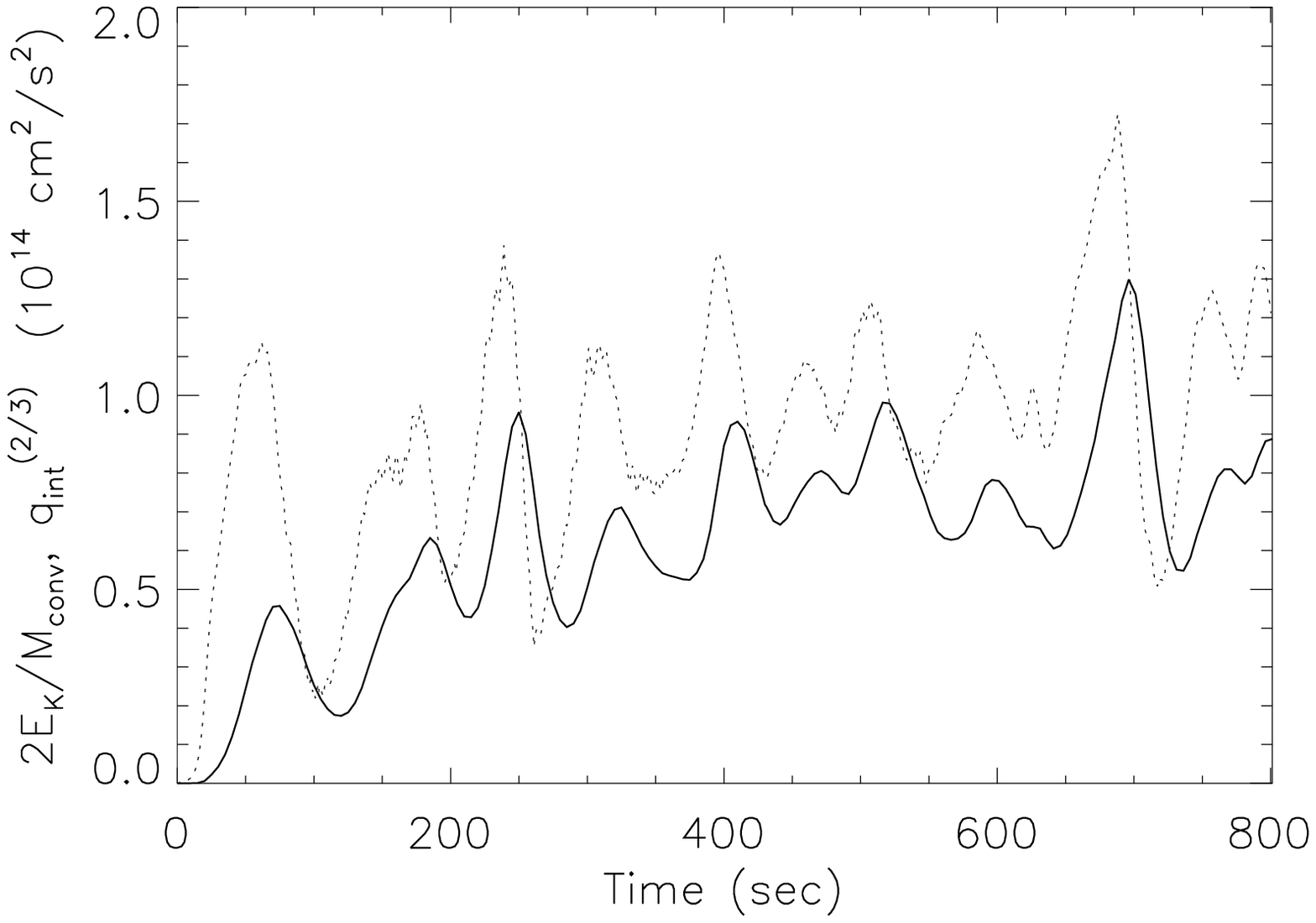}
\includegraphics[trim=550 -30 30 35,angle=-90,scale=0.25]{lor5.ps}
}
\caption{$\it Left.$ Turbulent velocity squared  $u_{t}^{2}=2 E_{turb}/M_{CZ}$(solid), and 
the corresponding buoyancy flux (dotted, plotted in the same units)
$q_{int}^{2/3}$  in the 
convection zone, versus time. 
The kinetic energy lags the buoyant flux by roughly 
20 seconds. 
$\it Right.$ Initial Behavior of an extended Lorenz Model of Convection,
for $\sigma=10$, $r=28$,  $b=8/3$, and 
$e = 3.0 \times 10^{-2} $. The parameters are similar to those
inferred for the oxygen shell (see \cite{am11}).
There is an increase in kinetic
energy and in the amplitude of the pulses, somewhat like the left panel.
The primary difference seems to be that the Lorenz model uses a single mode while
the simulation is multi-mode.
}
\label{fig1}
\end{figure}

\cite{lorenz} devised a simple model (three degrees of freedom)
 of a convective cell which captured the seeds of chaos in terms of the Lorenz 
 strange attractor, which is part of the foundation of the study of instabilities in
 nonlinear systems (\cite{cvit,gleick,thomp}). The model is an extreme truncation
 of the fluid dynamics equations, reducing a system of seven variables and about
 $10^7$ grid points (Fig.~1 (left)) to one of amplitudes of three variables
 (Fig.~1 (right)). This reduces the
 degrees of freedom from $\sim 7 \times 10^7$ to merely 3. To attain this simplification,
 only a single mode, of low Mach number flow, was examined in two dimensions
 (``a convective roll"). The three variables are the speed of the convective roll,
 the vertical temperature fluctuation, and the horizontal temperature fluctuation.
 For sufficiently large Reynolds number (essentially luminosity in excess of that
 which can be carried by radiative diffusion), the flow becomes chaotic.
 In MLT, the vertical and horizontal temperature fluctuations are assumed identical,
 and this two variable model is not chaotic (\cite{am11}).
 
 We examine the conjecture that the fluctuating behavior of the Lorenz model
 is a simple version of that seem in our 3D compressible, multi-mode convection simulations. 
 \cite{am11} make the simple generalization of the Lorenz model to include nuclear
 heating. In Fig.~1 is a comparison of the fluctuating behavior of kinetic energy for
 the 3D simulation (left panel) and the generalized Lorenz model (right panel).
 The behavior is similar, with the largest differences due to the fact that the 3D simulation
 exhibits many modes, while the Lorenz model has only one by definition.
 
 This suggests that the fluctuations seen in the 3D simulation are  related to the
 onset of chaotic behavior in the much simpler Lorenz model. Analytically, the
 equations of 3D fluid flow used in the simulations may be directly simplified to
 obtain the Lorenz equations (see \cite{am11}), which strengthens the suggestion.

\section{Convective Cells}

\begin{figure}
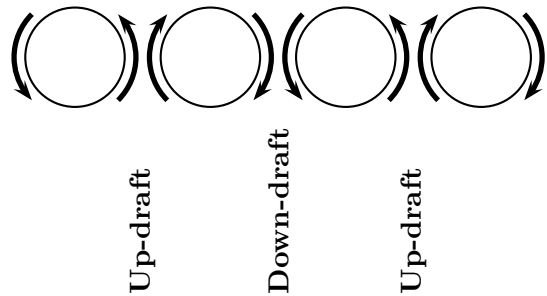

\center{
\
\psset{unit=.45cm}
\pspicture*[](-12,0)(4,10)
\psarc[linewidth=2pt]{->}(-10,8){1.8}{135}{225}
\pscircle[fillstyle=solid](-10,8){1.5}
\psarc[linewidth=2pt]{->}(-10,8){1.8}{-45}{45}

\psarc[linewidth=2pt]{<-}(-6,8){1.8}{135}{225}
\pscircle[fillstyle=solid](-6,8){1.5}
\psarc[linewidth=2pt]{<-}(-6,8){1.8}{-45}{45}

\psarc[linewidth=2pt]{->}(-2,8){1.8}{135}{225}
\pscircle[fillstyle=solid](-2,8){1.5}
\psarc[linewidth=2pt]{->}(-2,8){1.8}{-45}{45}

\psarc[linewidth=2pt]{<-}(2,8){1.8}{135}{225}
\pscircle[fillstyle=solid](2,8){1.5}
\psarc[linewidth=2pt]{<-}(2,8){1.8}{-45}{45}

\rput*[l]{L}(0,1){\large \bf Up-draft}
\rput*[l]{L}(-4,1){\large \bf Down-draft}
\rput*[l]{L}(-8,1){\large \bf Up-draft}

\endpspicture
}
\caption{The Lorenz Model extended: Convection in a shell composed of cells. Notice
the alternation of the sign of rotation. This may be thought of as a cross sectional view
of infinitely long cylindrical rolls, or of a set of toroidal cells, with pairwise alternating
vorticity (see \cite{ch61}, \S16). Each cell may exhibit independent fluctuations in time
and space.
}
\label{fig4}
\end{figure}
%\placefigure{4}

\cite{ms75} suggested, based upon MLT, that convection was dominated by cells,
whose size was determined by the size of a local pressure scale height. Simulations
suggest that this is qualitatively true; mass conservation in a stratified medium implies
cells of a density scale height in size (see \cite{nsa}).  Suppose that we imagine the
convective zone to be made of cells, each of which we approximate by a Lorenz model?
Fig.~2 illustrates a cross-section segment of part of such a planar convective layer.
We do not imagine that the cells are fixed as in a crystalline structure, but that the
cells are unstable, forming and decaying in a dynamic way, with average properties
approximated by independent Lorenz models.
Each Lorenz cell has a fluctuation in energy flux it carries, and this is identified with
a fluctuation in luminosity of that horizontal patch of the convective layer. For simplicity
we ignore the fluctuations of cells in underlying layers, and in the same spirit, the fact
that stellar photospheres often exhibit convective Mach numbers which are not small.

\cite{ms75} noted that for the sun, the local pressure scale height was relatively small,
implying that the size of a convective cell was also small relative to the radius, 
similar to the size of a granule.
Such a large number of cells ($\sim 10^6$) would average out fluctuations over a stellar
disk. For red supergiants, the scale heights approach  the radius in size, so that a few cells
would be sufficient to cover the surface, and the fluctuations in luminosity would be
more obvious.

\section{Irregular Variables}

\begin{figure}
\mbox{
\includegraphics[trim= 0 -10 30 35,angle=-90,scale=0.25]{fluc12.ps}
\includegraphics[trim=-50 480 30 35,scale=0.29]{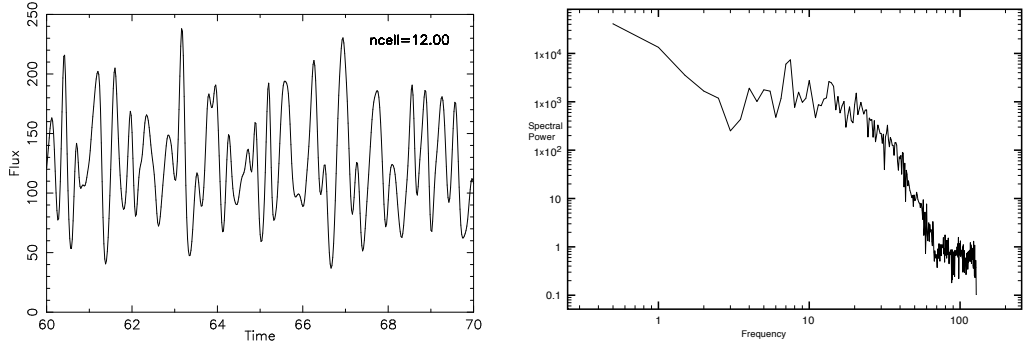}
}
\
\
\caption{(left) Fluctuations of Luminosity in Convective Layer of 12 Cells of Random Phase,
for $\sigma=10$, $r=28$, and $b=8/3$. 
The dimensionless flux (luminosity) is shown for a convective layer with 12 visible
Lorenz cells. The luminosity variations are large and seemingly chaotic, suggestive
of irregular variables and Betelgeuse in particular. (right) Power specta of the luminosity.
There is no sharp peak, but a broad distribution of power, as would be expected from
a chaotic source. 
}
\end{figure}

Combining Schwarzschild's idea of convective cells with Lorenz's model of a cell
allows us to construct a simple model of a stellar convective surface, which contains
an element of the fluctuating nature of turbulence, although it may to be
a crude approximation in that it has no realistic photospheric layer. 
Fig.~3 shows the luminosity fluctuations implied by such a model (left panel). The
amplitude of the luminosity fluctuations are simply those of the classical Lorenz 
model. They are larger than but comparable to those found in 3D simulations of
Betelgeuse ($\alpha$ Orionis) by \cite{chia10}, and observed in red supergiants
(\cite{kiss06}). This qualitative agreement for such a simple model is suggestive.

The right panel in Fig.~3 shows the spectral power density of the luminosity fluctuations
shown in the left panel. There are a few broad, marginal peaks, 
and a lot of broad-band noise.
The frequency is normalized to the time interval shown in the left panel.
In a real star, this broad-band noise would resonate with normal modes of pulsation, 
driving them to finite amplitude and thereby
giving additional sharp peaks to the spectrum.
 
Joel Stebbins (pioneer of photoelectric astronomy) 
monitored the brightness of Betelgeuse ($\alpha$ Orionis)
from 1917 to 1931, and concluded that ``there is no law or order in the rapid 
changes of Betelgeuse" (\cite{gold84}), which seems apt for Fig.~3 
as well. More modern observations (see \cite{kiss06}) show
a strong broad-band noise component in the photometric variability. 
The irregular fluctuations of the light curve are aperiodic, 
and resemble a series of outbursts.
This should be no surprise; the 3D equations have embedded in them
the strange attractor of Lorenz. It will be interesting to apply these ideas to
the analysis of solar-like oscillations in other stars (\cite{kb95}).

\section{Summary}
By quantitatively examining the implications of 3D simulations of turbulent convection,
we are led to an identification of similarities in convective cells and the Lorenz strange
attractor. Applying this connection in a simple model implies a broad-band
component of the power spectrum of luminosity in stars with convective surfaces,
much like that observed by \cite{kiss06}.

However, there is no free lunch. Application of these ideas to stellar surfaces implies
their use in regimes in which the convective Mach number $\mconv$ is not much less than unity. Will the compressible effects make qualitative changes in this picture? We shall see.
Fortunately the surface of the Sun is a nearby example on which we can test our theoretical ideas, and these ideas will have wide ramifications in astrophysics. 

\acknowledgements
This work was supported in part by by NSF Grant 0708871 and 
NASA Grant NNX08AH19G at the University of Arizona. 
We are grateful for helpful discussions with Aake Nordlund 
and Bob Stein, and thank the Symposium organizers for allowing us
to participate. Happy birthday Juri!

\begin{discussion}

\discuss{N. Weiss}{
A word of caution about the meaning of the Lorenz equations:
chaos is ubiquitous, and Lorenz's realisation is this was original and
important -- but one should not rely on his equations to describe a
specific physical system.
}
\discuss{W. D. Arnett}{I agree. We started with the simplest possibility, which
we knew contained chaotic behavior. 

The Lorenz model is only a 2D and incompressible model, and should be generalized to 3D, and compressible flow if possible. 
Hopefully, analysis of 3D simulations for a variety of conditions will illuminate 
this issue. 
}

\discuss{R. Collet}{
1. In your analysis you derive a value of the MLT parameter $\alpha$
that is four times as large as the typical value used in stellar
evolution calculations. This result is essentially based on
simulations of one massive star, if I understand correctly. How can
you generalise this result to other stellar masses, e.g solar?

2. Does this result apply to stellar interiors or to surface
convection as well?

}
\discuss{W. D. Arnett}{ 
Let me clarify. We claim that the $\alpha$ parameter is not adjustable, but an
eigenvalue determined by the stellar structure (depth of the convective zone).

Our original compressible 3D simulations (\cite{ma07b}) showed that 
the {\em turbulent dissipation length (essentially the ``mixing length") was simply equal to 
the depth of the convection zone} (2 pressure scale heights for that stellar model). Based on
this and
results from many 3D simulations by others, we conjectured that this relation was a general
result, but found that it seemed limited on the high side by a value of roughly 4 pressure scale
heights.  We have since verified the relation with 3D simulations of shells from 0.5 to 4 
pressure scale heights in depth for low Mach-number flows (stellar interiors). 
The solar convection zone is 20 pressure scale heights deep, so we used 4 as the limiting value, which is $4/1.65 \approx 2.4$ times the standard value of $\alpha$ for the same solar atmospheric model.

We believe this limit may be a general result for low Mach-number flows. For the outer, sub-photospheric layers in a stellar convective zone, the convective Mach-number rises,
and pressure perturbations become increasingly important. We do not yet have a more
general theory applicable to this small but important region. As a first step we simply see what 
happens if we ignore the effects of compressibility.
}

\discuss{R. Trapendach}{
How realistic in your simulation is the surface superadiabatic gradient when compared
to 3-D calculations. How to you model those layers?}

\discuss{W. D. Arnett}{ 
They are not realistic enough. We use the radiative diffusion approximation, but have not directly simulated an atmosphere. However, we have focused here on the possible
implications of chaotic behavior (turbulence) for stellar variability, with the simplest model
we know.

We are attempting to develop a general {\em theoretical understanding} of stellar turbulence.
It may be that our computational abilities now run ahead of our ability to assimilate what we
compute. How can we develop a mathematical (rather than numerical) understanding?
So far, we have a promising quantitative solution for low Mach-number 
convective zones (stellar interiors). 

Stellar atmospheres and immediate sub-surface regions
have convective Mach-numbers which increase toward unity. In such cases, compressible
fluid effects can no longer be neglected (shock, g- and p-mode wave generation, dynamic
expansion and contraction, to mention some of the most important). 

These effects also occur
in other interesting astrophysical situations: explosions of novae and supernovae, thermonuclear ignition flashes, thin-shell flashes, and vigorous radial pulsations, for example.

Many groups, including yours, have already solved this problem {\em numerically} for stellar
atmospheres.  We should
join forces in the quest of a theoretical model capable of reproducing the general features
of turbulent convection, which is not restricted to small Mach-number flow.
We have a developing theory and many of you have numerical solutions;  we
would be happy to work with you.
}

\end{discussion}

\end{document}